\documentclass[fleqn,10pt]{wlscirep}

\usepackage[utf8x]{inputenc}
\usepackage[T1]{fontenc}
\usepackage{multicol}
\usepackage{graphicx}
\usepackage{float}
\usepackage[format=plain]{caption}

\newcommand{\Te}{$^{130}$Te}
\newcommand{\Tl}{$^{208}$Tl}
\newcommand{\TeO}{TeO$_2$}
\newcommand{\enrLMO}{Li$_2^{100}$MoO$_4$}
\newcommand{\ovbb}{$0\nu\beta\beta$}
\newcommand{\onbb}{$0\nu\beta\beta$}

\newcommand{\Qbb}{$Q_{\beta\beta}$}
\newcommand{\ckky}{counts$/($keV$\cdot$kg$\cdot$yr$)$}
\newcommand{\G}{$\gamma$}

\newcommand{\A}{$\alpha$}

\title{Search for Majorana neutrinos exploiting millikelvin cryogenics with CUORE}

\author{The CUORE Collaboration\footnote{Lists of participants and their affiliations appear at the end of the paper.}}
\date{\today}

\begin{document}

\begin{abstract}
The possibility that neutrinos may be their own antiparticles, unique among the known fundamental particles, rises from the symmetric theory of
fermions proposed by E.Majorana in 1937 \citemaintext{Majorana:1937vz}.  
Given the profound consequences of such Majorana neutrinos, among them potentially explaining the matter-antimatter asymmetry of the universe via
leptogenesis\citemaintext{Fukugita:1986hr}, the Majorana nature of neutrinos commands intense experimental scrutiny globally and one of the primary experimental
probes is neutrinoless double beta (\onbb) decay. Here we show new results on the search for {\onbb} decay of $^{130}$Te, using the latest advanced cryogenic calorimeters
with the CUORE experiment\citemaintext{CUOREFirstResult}. CUORE, operating just 10mK above the absolute zero temperature, has pushed the state of the art on three frontiers:
the sheer mass held at such ultra-low temperatures, operational longevity, and the low levels of ionizing radiation emanating from the cryogenic infrastructure. 
We find no evidence for \onbb\ decay and set a lower bound of $T_{1/2}^{0\nu}>2.2\cdot10^{25}$ yr at a 90\% credibility interval. 
We discuss potential applications of the advances made with CUORE to other fields such as direct dark matter, neutrino and nuclear physics searches and large-scale quantum computing,
which can benefit from sustained operation of large payloads in a low-radioactivity, ultra-low temperature cryogenic environment. \\[10pt]
Published on: Nature 604, 53 (2022) \hfill DOI:~\href{https://doi.org/10.1038/s41586-022-04497-4}{10.1038/s41586-022-04497-4}

\end{abstract}

\maketitle
\begin{multicols*}{2}

\section*{Introduction}

The Standard Model (SM) of particle physics is a successful paradigm
for the number, properties and interactions of fundamental particles.
Nevertheless, the observation of neutrino oscillations indicates the incompleteness of the SM:
they imply non-vanishing neutrino masses, requiring an extension of the SM,
and violate three accidental symmetries connected to the flavor lepton numbers $L_e$, $L_{\mu}$ and $L_{\tau}$, leaving the difference between the baryon and lepton number,
$B-L$, as the only unprobed quantity.
A promising process to experimentally test $B-L$ is neutrinoless double beta (\onbb) decay,
in which a nucleus of mass number $A$ and charge $Z$ decays by the emission
of only two electrons: $(A,Z)\rightarrow(A,Z+2)+2e^-$.
We highlight that this process creates two electrons, namely two matter particles~\citemaintext{Vissani:2021gdw}.
This decay can be mediated by a variety of non-SM mechanisms
involving Majorana neutrino masses.
A minimal extension of the SM Lagrangian adds heavy Majorana neutrinos
that mix with the known neutrinos to produce a set of light Majorana neutrinos,
explaining the observed light neutrino masses~\citemaintext{Bilenky:2014uka}
and at the same time providing a mechanism to explain the baryon asymmetry in the universe~\citemaintext{Fukugita:1986hr}.
At this time, experimental searches for \onbb\ decay are the most sensitive means to corroborate this framework.

The \onbb\ decay signature is a peak in the spectrum of summed energy of the two emitted electrons at the mass difference (\Qbb) between the parent and daughter nuclei.
A worldwide quest is ongoing, involving a range of nuclei
such as $^{76}$Ge~\citemaintext{GERDA:PRL,Alvis:2019sil}, $^{136}$Xe~\citemaintext{Anton:2019wmi,KamLAND-Zen:2016pfg},
and \Te. The latter, in the form of \TeO\ cryogenic calorimeters,
is employed by the Cryogenic Underground Observatory for Rare Events, CUORE~\citemaintext{Arnaboldi:2002du,Artusa:2014lgv}.

To fully exploit the potential of \TeO\ crystals as cryogenic calorimeters,
the CUORE collaboration designed and built the largest dilution refrigerator ever constructed,
capable of cooling $\sim$1.5\,tonne of material to a temperature of $\sim$10\,mK
and maintaining it for years with a 90\% duty cycle ($1 \text{ t}=1,000$ kg).
In this Article, we describe the performance of CUORE over a 4\,yr measurement campaign
and the results of a new high-sensitivity \onbb\ decay search with over 1\,tonne$\cdot$yr of \TeO\ exposure. 

\section*{The CUORE experiment}

\begin{figure*}[!ht]
    \centering
    \includegraphics[width=1.\textwidth]{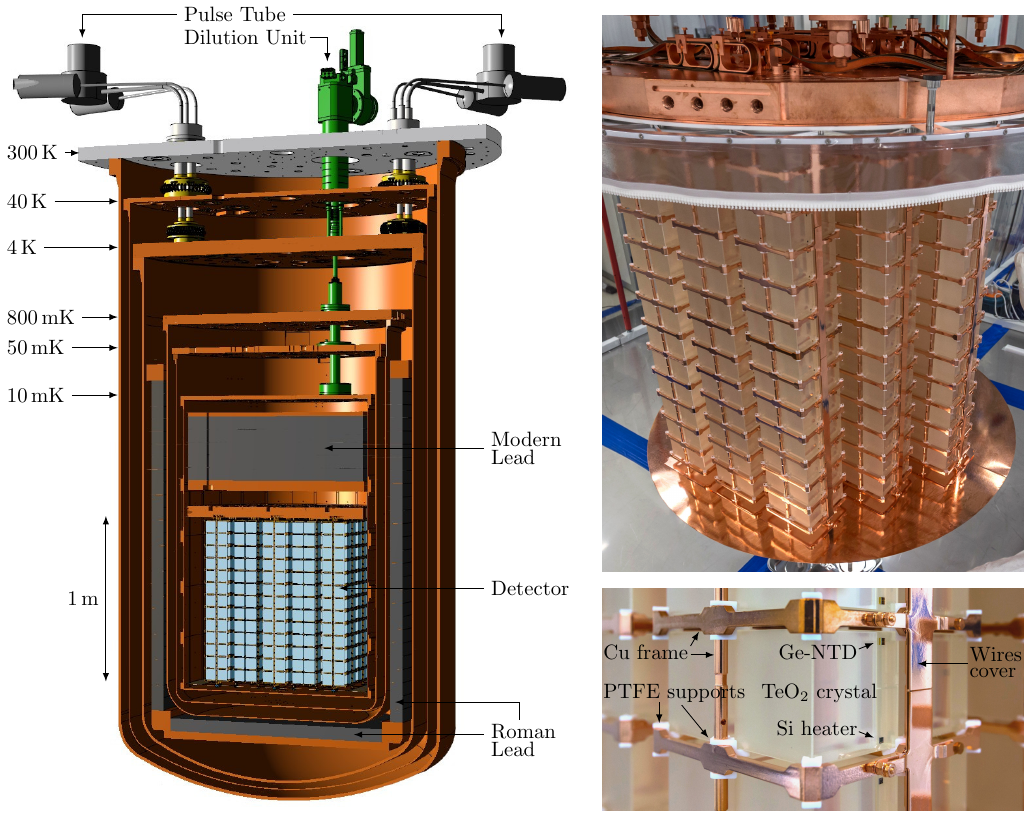}
    \caption{The CUORE detector. Left: Rendering of the 6-stage cryostat,
      with the pulse tubes and dilution unit,
      the internal low-radioactivity modern and Roman lead shields,
      and the array of 988 \TeO\ crystals (light blue). 
      Top right: The detector after installation.
      The plastic ring was used during assembly for radon protection.
      Bottom right: One of the calorimeters
      instrumented with an
      NTD Ge thermistor which measures the temperature increase induced by absorbed radiation.
      The Si heater is used to inject pulses for thermal gain stabilization.
      The polytetrafluorethylene (PTFE) supports and the gold wires instrumenting the NTD and the heater provide the thermal link between the crystal and the heat bath, i.e. the Cu frames~\protect\citemaintext{Alduino:2016vjd}.
    }
    \label{fig:detector}
\end{figure*}

CUORE is the culmination of thirty years of \onbb\ decay searches with \TeO\ cryogenic calorimeters~\citemaintext{Brofferio:2018lys}.
\Te\ benefits both from a high natural isotopic abundance of $\sim34\%$~\citemaintext{Fehr200483}
and a high \Qbb\ of $2527.5$\,keV~\citemaintext{Rahaman:2011zz},
placing the \onbb\ decay region of interest (ROI) above most natural \G-emitting radioactive backgrounds.
The detector is an array of 988 $^{\textrm{nat}}$\TeO\ cubic crystals~\citemaintext{Arnaboldi:2010fj} (Fig.~\ref{fig:detector})
of $5\times5\times5$\,cm$^3$ size and $\sim$750\,g mass, for a total mass of 742\,kg, which corresponds to 206\,kg of \Te.
The array is arranged as 19 towers, each comprised of 13 floors containing 4 crystals. 
The crystals are operated as cryogenic calorimeters~\citemaintext{Enss:2008} at a temperature of $\sim$10\,mK.  To achieve this low-temperature environment,
a novel cryogenic infrastructure --- the CUORE cryostat --- has been realized. 

In a cryogenic calorimeter, the energy deposited by impinging radiation in the absorber crystal is turned into heat, resulting in a temperature rise (Extended Data Fig. \ref{fig:CUORE_bolometer}).
Each CUORE crystal (Fig.~\ref{fig:detector}) is instrumented with a neutron-transmutation-doped
germanium thermistor (NTD)~\citemaintext{Haller1984} that converts thermal pulses into electric signals and a heater~\citemaintext{Andreotti:2012zz}
to inject reference heat pulses for thermal gain stabilization~\citemaintext{Carniti:2017zkr}.
Thermal signals can be induced by electrons emitted in \onbb\ decays but also other background radiation,
e.g. \G\ and \A\ particles from residual radioactive contaminants or cosmic ray muons.

CUORE is protected by several means against backgrounds that can mimic a \onbb\ decay.
It is located underground at the Laboratori Nazionali del Gran Sasso (LNGS) of INFN, Italy,
under a rock overburden equivalent to $\sim$3600\,m of water
which shields from hadronic cosmic rays and reduces the muon flux
by six orders of magnitude.
Environmental \G\ backgrounds are suppressed by a 30-cm layer of low-radioactivity lead above the detector (Fig.~\ref{fig:detector}),
a 6-cm-thick lateral and bottom shield of $^{210}$Pb-depleted ancient lead
recovered from a Roman shipwreck~\citemaintext{Pattavina:2019pxw} (Extended Data Fig~\ref{fig:roman_lead}),
and a 25-cm-thick lead shield outside the cryostat.
Environmental neutrons are suppressed by a 20-cm layer of polyethylene and a thin layer of boric acid outside of the external lead shield.
Finally, radioactive contaminants in the crystals and in the adjacent structures 
are minimized by careful screening of material for radio-purity and use of high-efficiency cleaning procedures
and manipulation protocols~\citemaintext{Alessandria:2013Cu}.

\begin{figure*}[!ht]
  \centering
  \includegraphics[width=1.\textwidth]{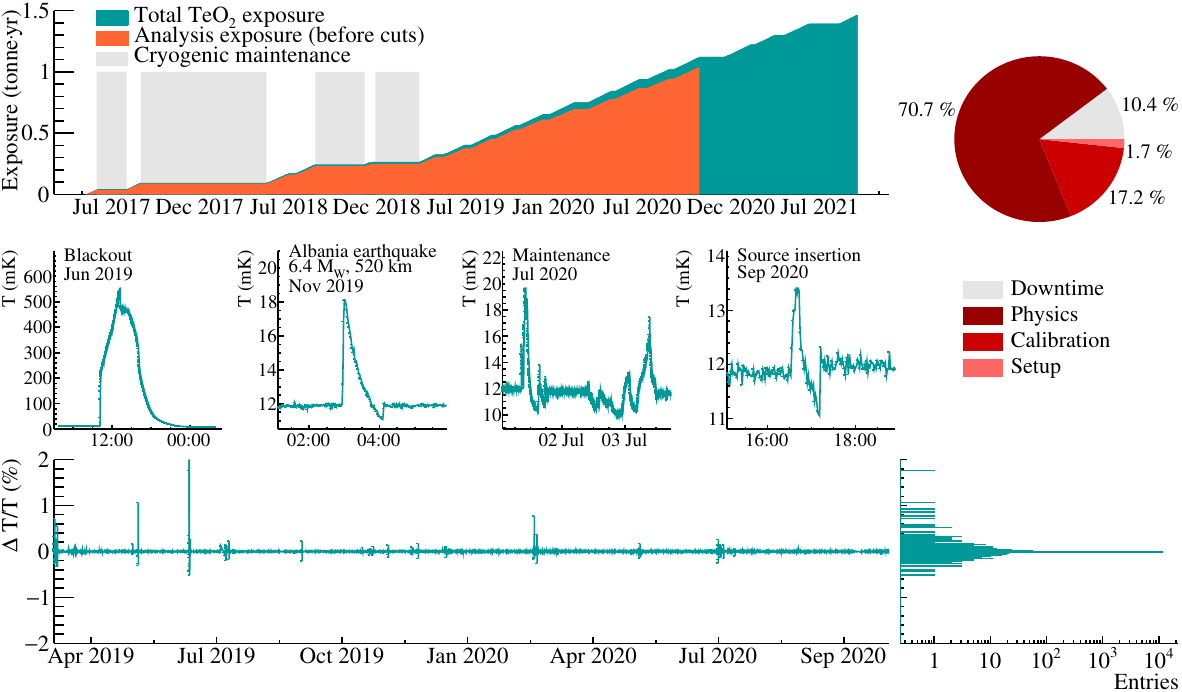}
  \caption{Cryogenic performance.
    Top: The exposure accumulated by CUORE (left, teal),
    along with the exposure used for this analysis (left, orange).
    Part of 2017 and 2018 was dedicated to maintenance and optimization of the cryogenic setup;
    since then, CUORE has been operating stably with a $90\%$ duty cycle (March 2019--October 2020) (right).
    Middle: Examples of temperature instabilities
    induced by external causes, e.g. blackouts and earthquakes,
    or human intervention, such as regular maintenance or the insertion of calibration sources.
    Bottom: The temperature stability of CUORE over $\sim$1\,yr of continuous operation, shown by a plot of relative temperature fluctuation versus time,
    and a histogram of the same data ($1 \text{ t yr} = 1,000 \text{ kg}\cdot\text{yr}$).
  }
  \label{fig:performance}
\end{figure*}

\section*{Cryogenic innovation and performance}

Dilution refrigerator technology was originally proposed in the '50s\citemaintext{London51} and underwent considerable development in the '80s
driven also by the application of cryogenic calorimeters for single particle detection~\citemaintext{CUORE:2021ctv}. Gradually, experimental volumes of the order of tens of
liters capable of hosting cold masses of up to 60 kg at 10~mK temperature~\citemaintext{Alduino:2016vjd} were achieved. Ultimately, detectors were limited by the capacity, duty cycle,
and radio-purity of commercial or near-commercial cryogenic systems. 
In the context of this history, the CUORE cryostat represents a breakthrough in cryogenic technology,
reaching an experimental volume of $\sim$1 m$^3$ and a cold mass of 1.5 tonne (detectors, holders, shields) at 10\,mK,
which correspond to a 20-fold improvement in experimental volume and target mass
compared to the previous state of the art at this temperature scale.
Prior to CUORE, the ultimate temperature for comparable target masses was in the resonant-mass gravitational antenna community at 65 mK~\citemaintext{CUORE:2021ctv}.

The CUORE detector is hosted in a multistage cryogen-free cryostat~\citemaintext{Alduino:2019xia} (Fig.~\ref{fig:detector}),
equipped with 5 pulse tube (PT) cryocoolers that avoid pre-cooling with a liquid helium bath thus enabling a high duty-cycle.
A custom-designed dilution unit with a double condensing line for redundancy
provides more than 4\,$\mu$W cooling power at 10\,mK. The cryostat is uniquely designed to provide
the necessary i) cooling power and temperature stability over a time scale of years, ii) very low radioactivity environment, and iii) extremely low vibration conditions.
As shown in Fig.~\ref{fig:performance} (top and right),
CUORE became operational in 2017, with the initial period mostly devoted to characterization and optimization campaigns.
Since 2019, the data taking has proceeded smoothly with a duty cycle of $\sim90\%$.
Fig. \ref{fig:performance} (bottom) shows that the temperature stability achieved is at the level of $0.2\%$ ($\pm3\sigma$ range) over a period in excess of 1 year.
Such a stability is important to achieve a uniform response of all detectors over time.
The CUORE calorimeters are sensitive to thermal signals and feature an intrinsic thermal fluctuation limit of $\sim0.5$\,keV,
so any process inducing heat dissipation $\geq0.5$\,keV degrades the energy resolution.
Mechanical vibrations can be transferred to the inner components and produce heat through friction.
To minimize the impact of vibrational noise, the calorimeter array is mechanically decoupled from the cryostat by a custom suspension system.
Vibrations induced by the PTs at the 1.4\,Hz operational frequency and its harmonics are particularly relevant.
In CUORE, we actively tune the PT relative phases for vibration cancellation~\citemaintext{DAddabbo:2017efe} (Fig. \ref{fig:ptphases}).
This solution is transferable to any cryogenic application involving signals in the same bandwidth of the PT-induced noise. 

CUORE now collects sensitive exposure with 984 out of 988 calorimeters,
at a rate which is, to our knowledge, unprecedented for cryogenic particle detectors.
Below, we describe the data treatment and \onbb\ decay search with  $>$1\,tonne$\cdot$yr of \TeO\ exposure.

\begin{figure}[H]
  \centering
  \caption{PT phase optimization. Top: frequency spectrum of the noise for a random combination
    of the PT phases (orange) and after the active phase tuning (teal).
    The frequency spectrum of physical signals is also reported for reference.
    Bottom: integral of the power spectrum at the PT frequency (1.4\,Hz) and its harmonics
    before and after active noise cancellation.}
  \label{fig:ptphases}
\end{figure}

\section*{Data Analysis and Results}
CUORE data are subdivided into datasets of 1-2 months of \emph{physics data}, separated
by a few days of calibration data collected with the detector exposed to $^{232}$Th and/or $^{60}$Co sources. 

The voltage across each NTD is amplified, passed through 
an anti-aliasing filter, and continuously digitized with a 1\,kHz sampling frequency~\citemaintext{Arnaboldi:2017aek,DiDomizio:2018ldc}.
We identify thermal pulses in the data stream and compute the pulse amplitudes,
applying optimum filters that maximize the frequency-dependent
signal-to-noise ratio~\citemaintext{DiDomizio:2010ph}.
To monitor and correct for possible drifts of the thermal gain of the detectors we exploit two {\emph{standard candles}}: monoenergetic heater pulses for the calorimeters
with functioning and stable heaters ($>95\%$ of the total), and events from the 2615\,keV \Tl\ calibration line for the remainder.  Drift-stabilized pulse amplitudes are
converted to energy using the regularly acquired source calibration data~\citemaintext{CUORE0AnalysisTechniques}.
We blind the {\onbb} search via a data salting procedure
that produces an artificial peak at \Qbb ~\citemaintext{CUORE0AnalysisTechniques}. Once the full analysis procedure is finalized and frozen, we reverse the salting.

To simplify the analysis, we eliminate data from periods impacted by high noise or failed data processing, which amounts to 5\% of the exposure. Furthermore calorimeters with
$>19$\,keV full width at half maximum (FWHM) energy resolution
at the 2615\,keV calibration line are discarded, adding 3\% loss in exposure.
In addition to these so-called base cuts, the following second-level cuts are then applied to suppress single background-like or low-quality events. Monte Carlo (MC) simulations
show that $\sim88\%$  of \onbb\ decay events
release their full energy in a single crystal~\citemaintext{CUORE:2020bok}.
Hence, we apply an anti-coincidence (AC) cut that excludes events depositing energy in more than one crystal.
Finally, we use pulse shape discrimination (PSD) to eliminate pulses consistent with more than one energy deposit in the pulse time window, pulses with a non-physical shape,
and excessively noisy pulses that survived the previous selections (Extended Data Fig. \ref{fig:PSDcuts}).
The selection efficiencies are summarized in Tab.~\ref{tab:Efficiencies}, with details provided in Methods.

The detector response to a monoenergetic energy deposition is an important input to the {\onbb} decay search. We empirically model the response function of each calorimeter 
as a sum of three equal-width Gaussians and determine the function parameters from a fit to the 2615\,keV calibration line ~\citemaintext{CUOREFirstResult}. As a characteristic 
indicator of the overall energy resolution of the calorimeters we quote the exposure-weighted harmonic mean FWHM of the detectors at this calibration line, $(7.78\pm0.03)$\,keV.
All values are reported as \mbox{mean $\pm$ s.\,d.\,.}

We quantify the scaling of energy resolution with energy and investigate energy reconstruction bias, i.e. the deviation of reconstructed {\G}-line positions from the literature values,
by fitting the calorimeter response functions to prominent \G\ lines in the physics data, allowing the peak means and widths to vary in the fit.  The bias is allowed to scale as a quadratic
function of energy as done in our previous result \citemaintext{CUOREPRLResult}, while the resolution scaling has been changed to a linear function of energy, following studies showing that
it was overparameterized by a quadratic scaling.  The results, extrapolated to \Qbb, are an exposure-weighted harmonic mean FWHM energy resolution 
of $(7.8\pm0.5)$\,keV and an energy bias of $<0.7$\,keV.  We summarise all the relevant analysis quantities in Tab.~\ref{tab:Efficiencies}.

\begin{figure*}[!ht]
  \centering
  \includegraphics[width=1.\textwidth]{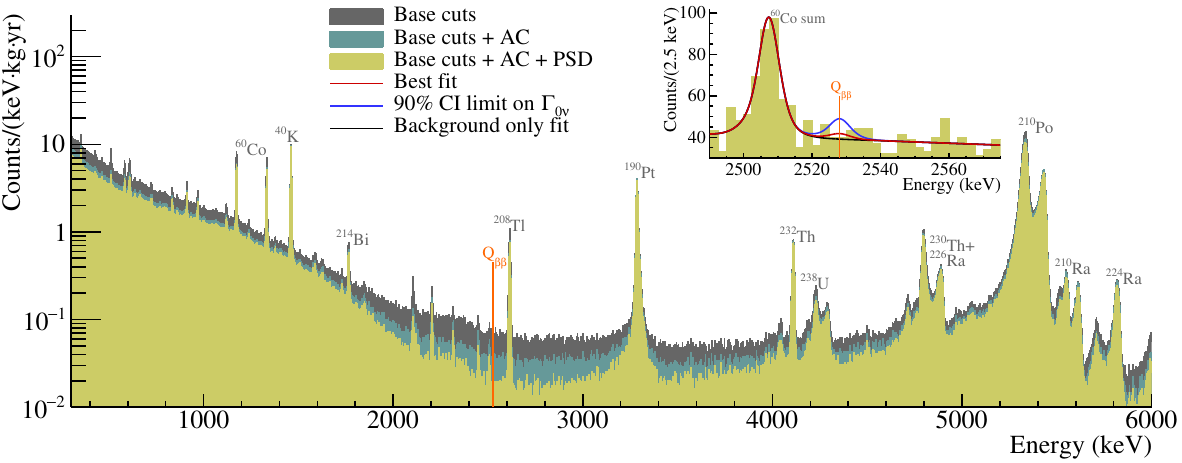}
  \caption{Physics spectrum for 1038.4\,kg$\cdot$yr of \TeO\ exposure.
    We separately show the effects of the base cuts, the anti-coincidence (AC) cut,
    and the pulse shape discrimination (PSD). The most prominent background peaks in the spectrum are highlighted.
    Top right inset: the ROI after all selection cuts, with the best-fit curve (solid red),
    the best-fit curve with the $0\nu\beta\beta$ rate fixed to the 90\% CI limit (blue),
    and background-only fit (black) superimposed.}
  \label{fig:BkgSpectrum}
\end{figure*}

\begin{table}[H]
  \centering
  \caption{Main parameters for the {\onbb} analysis.
    The resolution and efficiencies are exposure-weighted average values.}\label{tab:Efficiencies}
  \begin{tabular}{ll}
    \toprule
    \midrule
    Number of datasets &  15 \\
    TeO$_2$ exposure & 1038.4 kg$ \cdot$ yr\vspace{1pt}\\
    $^{130}$Te exposure & 288.8 kg $\cdot$ yr\vspace{1pt}\\
    \hline\vspace{-10pt} \\
    FWHM at 2615 keV in calibration data & 7.78(3) keV\\
    FWHM at \Qbb\ in physics data & 7.8(5) keV\vspace{1pt}\\ \hline\vspace{-10pt} \\
    Total analysis efficiency (data) & 92.4(2)\%\\
    \quad Reconstruction efficiency & 96.418(2)\%\\
    \quad Anticoincidence efficiency & 99.3(1)\%\\
    \quad PSD efficiency & 96.4(2)\%\\
    Containment efficiency (MC) & 88.35(9)\%~\citemaintext{CUORE0AnalysisTechniques} \\
    \midrule
    \bottomrule
  \end{tabular}   
\end{table}

Fig.~\ref{fig:BkgSpectrum} shows the full energy spectrum along with the $[2490,2575]$\, keV fit region,
which contains only one background peak at 2505.7\,keV from the simultaneous absorption of two coincident \G\ rays from $^{60}$Co in the same crystal. We estimate $\sim$90\% of the
continuum background consists of degraded \A\ particles from
radioactive contaminants of the support structure surface, while the other $\sim$10\% are multi-Compton scattered 2615\,keV \G\ events~\citemaintext{Alduino:2017qet,CUORE:2020bok}.

We run an unbinned Bayesian fit with uniform non-negative priors on the background and \onbb\ decay rates.
The fit with a background-only model, i.e. excluding the \onbb\ component,
yields a mean background rate of $(1.49\pm0.04)\cdot10^{-2}$\,\ckky\ at \Qbb\
for a corresponding median exclusion sensitivity of $T^{0\nu}_{1/2}>2.8\cdot 10^{25}$\,yr (90\% credibility interval (CI)).
The fit with the signal-plus-background model shows no evidence for \onbb\ decay.
We find the best fit at \mbox{$\Gamma_{0\nu}=(0.9\pm1.4)\cdot10^{-26}$\,yr$^{-1}$}
and set a limit on the process half-life
of $T^{0\nu}_{1/2} > 2.2\cdot10^{25}$\,yr (90\% CI).
Systematic uncertainties are included as nuisance parameters and affect both the best fit and the limit by 0.8\% (Extended Data Tab.~\ref{tab:Systematics}).
Compared to the sensitivity, the probability of getting a stronger limit is 72\%.
This represents, to our knowledge, the current world-leading $0\nu\beta\beta$ sensitivity for \Te, having improved in accordance with our increased exposure since our previous result\cite{CUOREPRLResult},
and although the actual limit is weaker, it is well within the range of possible outcomes due to statistical fluctuations.

Under the common assumption of a light neutrino exchange mechanism,  the \Te\ half-life limit
converts to a limit on the effective Majorana mass of $m_{\beta\beta}<90$-305\,meV,
with the spread induced by different nuclear matrix element
calculations~\citemaintext{Menendez:2008jp,Simkovic:2013qiy,Vaquero:2014dna,Neacsu:2014bia,Yao:2014uta,Barea:2015kwa,Hyvarinen:2015bda}.
This limit on $m_{\beta\beta}$ is among the strongest in the field,
proving the competitiveness of the cryogenic calorimeter technique used in CUORE.
CUORE will continue to take data until it reaches its designed \Te\ exposure of 1000 kg$\cdot$yr.

\section*{Impact}

We have demonstrated for the first time that the cryogenic calorimeter technique is scalable to tonne-scale detector masses and multi-year measurement campaigns, while maintaining low
radioactive backgrounds. Next generation calorimetric \onbb\ decay searches exploiting these developments are planned. Among these, CUPID (CUORE Upgrade with Particle
IDentification)~\citemaintext{CUPIDInterestGroup:2019inu}
will utilize the same cryogenic infrastructure as CUORE,
replacing the \TeO\ crystals with scintillating \enrLMO\ crystals
and exploiting the scintillation light for $>100$-fold active suppression of the \A\ background~\citemaintext{Armengaud:2019loe,CUPID:2019gpc}.
In parallel, the AMoRE collaboration aims to build a large mass calorimetric \onbb\ decay experiment in Korea \citemaintext{AMORE:2019}. 
In general, the possibility to cool large detector payloads paired with the low energy thresholds achievable by cryogenic calorimeters
will benefit next-generation projects at the frontier of particle physics, such as dark matter searches
like SuperCDMS~\citemaintext{SuperCDMS:2016wui} and CRESST~\citemaintext{CRESST:2019jnq}, and low-energy observatories exploiting CE$\nu$NS
for solar and supernova neutrino studies~\citemaintext{Pattavina:2020cqc} and neutrino flux monitoring of nuclear reactors~\citemaintext{nucleus2019}. 

A quite serendipitous impact is that the cryogenic innovations pioneered by CUORE for {\onbb}  decay appear to be a solution-in-waiting for the challenges faced by the relatively young,
but rapidly growing field of quantum information technology. The need to cool increasingly large arrays of qubits to $\lesssim100$\,mK means there is now a commercial market for large,
high-cooling-power dilution refrigerators with some featuring technological solutions derived from CUORE. 
Moreover, the recent realization that ionizing radiation from natural radioactivity will be a limiting factor for the coherence time of quantum processors with an increasing number of
qubits~\citemaintext{Wilen:2020lgg}
suggests the one-time niche, low-radioactivity ultra-low temperature cryogenics pioneered for {\onbb} decay may become mainstream in large-scale quantum computing~\citemaintext{Cardani_2021}. 

\section*{Online Content}
Methods, additional Extended Data and Source Data are available in the online version of the paper.



\section*{Acknowledgments}
The CUORE Collaboration thanks the directors and staff of the Laboratori Nazionali del Gran Sasso
and the technical staff of our laboratories.
This work was supported by the Istituto Nazionale di Fisica Nucleare (INFN);
the National Science Foundation under Grant Nos. NSF-PHY-0605119, NSF-PHY-0500337,
NSF-PHY-0855314, NSF-PHY-0902171, NSF-PHY-0969852, NSF-PHY-1614611, NSF-PHY-1307204,
NSF-PHY-1314881, NSF-PHY-1401832, and NSF-PHY-1913374; and Yale University.
This material is also based upon work supported by the US Department of Energy (DOE)
Office of Science under Contract Nos. DE-AC02-05CH11231 and DE-AC52-07NA27344;
by the DOE Office of Science, Office of Nuclear Physics under Contract Nos.
DE-FG02-08ER41551, DE-FG03-00ER41138, DE- SC0012654, DE-SC0020423, DE-SC0019316;
and by the EU Horizon2020 research and innovation program under
the Marie Sklodowska-Curie Grant Agreement No. 754496.
This research used resources of the National Energy Research Scientific Computing Center (NERSC).
This work makes use of both the DIANA data analysis and APOLLO data acquisition software packages,
which were developed by the CUORICINO, CUORE, LUCIFER and CUPID-0 Collaborations.

\section*{Author Contributions}
All listed authors have contributed to the present publication. The different contributions span from the design and construction of the detector and of the cryogenic system to the
acquisition and analysis of data. 
The manuscript underwent an internal review process extended to the whole Collaboration and all authors approved its final version; the authors' names are listed alphabetically.

\section*{Author Information}
Reprints and permissions information is available at www.nature.com/reprints.
The authors declare no competing interests.
Correspondence and requests for materials should be addressed to the CUORE Collaboration (cuore-spokesperson@lngs.infn.it).

\section*{The CUORE collaboration}
\noindent 
D.~Q.~Adams$^{1}$, C.~Alduino$^{1}$, K.~Alfonso$^{2}$, F.~T.~Avignone~III$^{1}$, O.~Azzolini$^{3}$,
G.~Bari$^{4}$, F.~Bellini$^{5,6}$, G.~Benato$^{7}$, M.~Beretta$^{8}$, M.~Biassoni$^{9}$
% ...

\vspace{4mm}

\noindent 
$^{1}$ Department of Physics and Astronomy, University of South Carolina, Columbia, SC 29208, USA \\
$^{2}$ Department of Physics and Astronomy, University of California, Los Angeles, CA 90095, USA \\
$^{3}$ INFN -- Laboratori Nazionali di Legnaro, Legnaro (Padova) I-35020, Italy \\
$^{4}$ INFN -- Sezione di Bologna, Bologna I-40127, Italy \\
$^{5}$ Dipartimento di Fisica, Sapienza Universit\`{a} di Roma, Roma I-00185, Italy \\
$^{6}$ INFN -- Sezione di Roma, Roma I-00185, Italy \\
$^{7}$ INFN -- Laboratori Nazionali del Gran Sasso, Assergi (L'Aquila) I-67100, Italy \\
$^{8}$ Department of Physics, University of California, Berkeley, CA 94720, USA \\
$^{9}$ INFN -- Sezione di Milano Bicocca, Milano I-20126, Italy \\
% ...

\section*{Methods}
\subsection*{Optimum trigger and analysis threshold}
\label{subsec:OT}

The continuous data stream of CUORE is triggered with the optimum trigger (OT),
an algorithm based on the optimum filter~\citemethods{Gatti:1986cw}
characterized by a lower threshold than a more standard derivative trigger~\citemaintext{CUOREPRLResult}.
A lower threshold allows us not only to reconstruct the low-energy part of the spectrum,
but also yields a higher efficiency in reconstructing the events in coincidence
between different calorimeters, and consequently a more precise understanding
of the corresponding background components~\citemethods{Alduino:2017xpk,Campani:2020ltd}.

The OT transfer function of every event is matched to the ideal signal shape,
obtained as the average of good quality pulses, so that frequency components
with low signal-to-noise ratio are suppressed.
A trigger is fired if the filtered signal amplitude exceeds a fixed multiple
of the noise root mean square (RMS), defined separately for each calorimeter and dataset.
We evaluate the energy threshold by injecting fake pulses of varying amplitude,
calculated by inverting the calibration function, into the data stream.
We reconstruct the stabilized amplitude of the fake pulses,
fit the ratio of correctly triggered events to generated events with an error function,
and use the 90\% quantile as a figure of merit for the OT threshold.
This approach allows monitoring of the threshold during data collection,
and is based on the assumption that the signal shape is not energy dependent,
i.e. that the average pulse obtained from high energy \G\ events
is a good template also for events of a few keV.
The distribution of energy threshold at 90\% trigger efficiency is shown in Extended Data Fig.~\ref{fig:OT_thresholds}.

For this work we set a common analysis threshold of 40\,keV
which results in $>90\%$ trigger efficiency for the majority ($97\%$) of the calorimeters
while at the same time allowing the removal of multi-Compton events from the ROI through the AC cut.

\subsection*{Efficiencies}

The total efficiency is the product of the reconstruction,
AC, PSD and containment efficiencies.

The reconstruction efficiency is the probability that a signal event is triggered,
has the energy properly reconstructed, and is not rejected by the basic quality cuts
requiring a stable pre-trigger voltage and only a single pulse in the signal window.
It is evaluated for each calorimeter using externally flagged heater events~\citemethods{Alessandrello:1998bf},
which are a good approximation of signal-like events.

The AC efficiency is the probability that a true single-crystal event
correctly passes our AC cut, instead of being wrongly vetoed due to an accidental coincidence with an unrelated event.
It is extracted as the acceptance of fully absorbed \G\ events at 1460\,keV from the electron capture decays of $^{40}$K,
which provide a reference sample of single-crystal events.

The PSD efficiency is obtained as the average acceptance of events in the $^{60}$Co, $^{40}$K,
and \Tl\ \G\ peaks that already passed the base and anti-coincidence cuts. 
In principle, the PSD efficiency could be different for each calorimeter,
but given the limited statistics in physics data we evaluate it over all channels and over the full dataset. 
To account for possible variation between individual calorimeters,
we compare the PSD efficiency obtained by directly summing their individual spectra
with that extracted from an exposure-weighted sum of the calorimeters' spectra.
We find an average $\pm0.3\%$ discrepancy between the two values
and include it as a global systematic uncertainty in the \ovbb\ fit. This takes a Gaussian prior instead of the uniform prior used in our previous
result\cite{CUOREPRLResult}, which had its uncertainty come from a discrepancy between two approaches that has since been resolved.

Finally, the containment efficiency is evaluated through Geant4-based MC simulations~\citemethods{CUORE:2016ons}
and accounts for the energy loss due to geometrical effects as well as bremsstrahlung.

\subsection*{Principal Component Analysis for PSD}

In this analysis we use a new algorithm based on principal component analysis (PCA) for pulse shape discrimination.
The method has been developed and documented for CUPID-Mo~\citemethods{Huang:2020mko},
and has been adapted for use in CUORE\citemethods{RHuangThesis}.
This technique replaces the algorithm employed in previous CUORE results,
which was based on 6 pulse shape variables~\citemaintext{CUORE0AnalysisTechniques}.
The PCA decomposition of signal-like events
pulled from $\gamma$ calibration peaks
yields a leading component similar to an average pulse,
which on its own captures $>90\%$ of the variance between pulses.
We choose to treat the average pulse of each calorimeter in a dataset
as if it were the leading PCA component, normalizing it like a PCA eigenvector.
We can then project any event from the same channel onto this vector
and attempt to reconstruct the 10-second waveform using only this leading component.
For any waveform $\mathbf{x}$ and leading PCA component $\mathbf{w}$ with length $n$,
we define the \textit{reconstruction error} as:

\begin{equation}
  RE=\sqrt{\sum\limits_{i=1}^n \left(\mathbf{x}_i - (\mathbf{x}\cdot\mathbf{w})\mathbf{w}_i\right)^2}
\end{equation}

This reconstruction error metric measures how well an event waveform
can be reconstructed using only the average pulse treated as a leading PCA component.
Events that deviate from the typical expected shape of a signal waveform
are poorly reconstructed and have a high reconstruction error.
We normalize the reconstruction errors as a second order polynomial function of energy
on a calorimeter-dataset basis (see Extended Data Fig. \ref{fig:pca_normfit}),
and cut on the normalized values by optimizing a figure of merit
for signal efficiency over expected background in the \Qbb\ ROI.
Using this PCA-based method, we obtain an overall efficiency of $(96.4\pm0.2)\%$
compared to the $(94.0\pm0.2)\%$ from the pulse shape analysis used in our previous results,
as well as a 50\% reduction in the PSD systematic uncertainty from 0.6\% to 0.3\%.

\subsection*{Statistical analysis}

The high-level statistical \onbb\ decay analysis consists of
an unbinned Bayesian fit on the combined data developed with the BAT software package \citemethods{BATSoftware}.
The model parameters are the \onbb\ decay rate ($\Gamma_{0\nu}$),
a linearly sloped background, and the $^{60}$Co sum peak amplitude.
$\Gamma_{0\nu}$ and the $^{60}$Co rate are common to all datasets,
with the $^{60}$Co rate scaled by a preset dataset-dependent factor
to account for its expected decay over time. The base background rate,
expressed in terms of \ckky, is dataset-dependent,
while the linear slope to the background is shared among all datasets
since any structure to the shape of the background should not vary between datasets.
$\Gamma_{0\nu}$, the $^{60}$Co rate, and the background rate parameters
have uniform priors constrained to non-negative values,
while the linear slope to the background has a uniform prior allowing both positive and negative values.

In addition to these statistical parameters, we consider the systematic effects
induced by the uncertainty on the energy bias and energy resolution\citemethods{GFantiniThesis,ACampaniThesis},
the value of \Qbb, the natural isotopic abundance of \Te,
and the reconstruction, AC, PSD and containment efficiencies.
We evaluate their separate effects on the \ovbb\ rate
by adding nuisance parameters to the fit one at a time
and studying both the effect on the posterior global mode $\hat{\Gamma}_{0 \nu}$
and the marginalized 90\% CI limit on $\Gamma_{0\nu}$.

A list of the systematics and priors is reported in Extended Data Tab.~\ref{tab:Systematics}.
The efficiencies and the isotopic abundance are multiplicative terms on our expected signal,
so the impact of each is reported as a relative effect on $\Gamma_{0\nu}$.
In contrast, the uncertainties on \Qbb, the energy bias,
and the resolution scaling have a non-trivial relation to the signal rate;
therefore, we report the absolute effect of each on $\Gamma_{0\nu}$. 
The dominant effect is due to the uncertainty on the energy bias
and resolution scaling in physics data. We account for possible correlations
between the nuisance parameters by including all of them in the fit simultaneously.

We chose a uniform prior on our physical observable of interest $\Gamma_{0\nu}$,
which means we treat any number of signal events as equally likely.
Other possible uninformative choices could be considered appropriate, as well.
Since the result of any Bayesian analysis depends to some extent on the choice of the priors,
we checked how other reasonable priors affect our result\citemethods{RHuangThesis}.
We considered:
\begin{itemize}
\item a uniform prior on $\sqrt{\Gamma_{0\nu}}$, equivalent to a uniform prior on $m_{\beta\beta}$ and also equivalent to using the Jeffreys prior;
\item a scale-invariant uniform prior on $\log{\Gamma_{0\nu}}$;
\item a uniform prior on $1/\Gamma_{0\nu}$, equivalent to a uniform prior on $T^{0\nu}_{1/2}$.
\end{itemize}
These priors are all undefined at $\Gamma_{0\nu}=0$,
so we impose a lower cut-off of $\Gamma_{0\nu}>10^{-27}$\,yr$^{-1}$,
which with the given exposure corresponds to approximately one signal event.
The case with a uniform prior on $\sqrt{\Gamma_{0\nu}}$ gives the smallest effect,
and strengthens the limit by 25\%,
while the flat prior on $1/\Gamma_{0\nu}$ provides the largest effect, increasing the limit on $T_{1/2}^{0\nu}$ by a factor of 4.
In fact, all these priors weigh the small values of $\Gamma_{0\nu}$ more.
Therefore, our choice of a flat prior on $\Gamma_{0\nu}$ provides the most conservative result.

We compute the \onbb\ exclusion sensitivity by 
generating a set of $10^4$ toy experiments with the background-model,
i.e. including only the $^{60}$Co and linear background components.
The toys are split into 15 datasets with exposure and background rates
obtained from the background-only fits to our actual data.
We fit each toy with the signal-plus-background model,
and extract the distribution of 90\% CI limits,
shown in Extended Data Fig. \ref{fig:Sensitivity}.

We perform the frequentist analysis using the Rolke method \citemethods{Rolke2001745},
obtaining a lower limit on the process half-life of $T_{1/2}^{0\nu}>2.6\cdot10^{25}$ yr ($90\%$ CI).
The profile likelihood function $\mathcal{L}$ for $\Gamma_{0\nu}$
is retrieved from the full Markov Chain produced by the Bayesian analysis tool.
The non-uniform priors on the systematic effects in the Bayesian fit are thus incorporated into the frequentist result as well.
We extract a 90\% confidence interval on $\Gamma_{0\nu}$
by treating $-2\log\mathcal{L}$ as an approximate $\chi^2$ distribution with one degree of freedom.
The lower limit on $T_{1/2}^{0\nu}$ comes from the corresponding upper edge of the confidence interval on $\Gamma_{0\nu}$.
Applying the same method to the toy experiments,
we find a median exclusion sensitivity of $T^{0\nu}_{1/2}>2.9\cdot10^{25}$ yr.

\section*{Data Availability}
The data generated during this analysis and shown in paper figures are available in ASCII format (CSV) as Source Data in the repository
\emph{https://cuore.lngs.infn.it/en/publications/collaborationpapers}. Additional information is available upon request by contacting the CUORE Collaboration.


\end{multicols*}

\renewcommand{\figurename}{Extended Data Figure}
\setcounter{figure}{0}
\renewcommand{\tablename}{Extended Data Table}
\setcounter{table}{0}
\begin{figure}[tb]
  \centering
    \includegraphics[width=1.\textwidth]{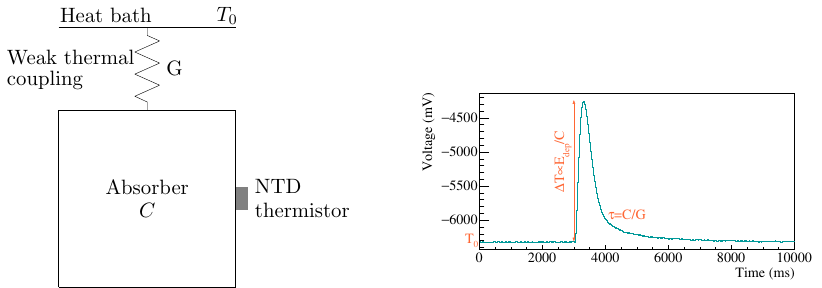}
    \caption{Working principle of the cryogenic calorimeter. Left: simplified calorimeter thermal model.
    The detector is modeled as a single object with heat capacity $C$
    coupled to the heat bath (with constant temperature $T_0$) through the thermal conductance $G$.
    The NTD thermistor for signal readout is glued to the absorber.
    Right: Example of a CUORE pulse from the 2615\,keV calibration line:
    $T_0$ corresponds to the baseline height,
    the pulse amplitude is proportional to the deposited energy,
    and the decay time depends on the $C/G$ ratio.}
  \label{fig:CUORE_bolometer}
\end{figure}

\begin{figure}
  \centering
  \includegraphics[width=1.\textwidth]{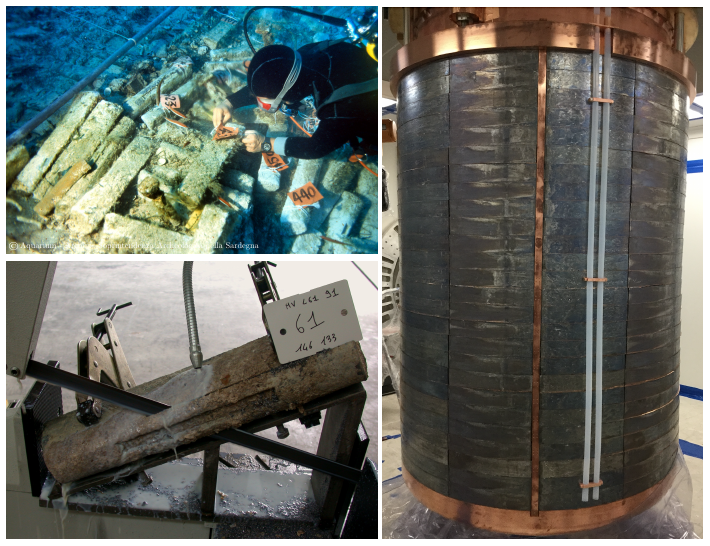}
  \caption{Roman lead. Top left: the lead bricks recovery from the Sardinian sea.
    Bottom left: the ingot inscriptions were cut and preserved,
    while the ingot bodies were used for the CUORE internal lead shield.
    Right: Lateral view of the internal lead shield~\protect\citemaintext{Pattavina:2019pxw}.}
  \label{fig:roman_lead}
\end{figure}

\begin{figure}
  \centering
  \includegraphics[width=\textwidth]{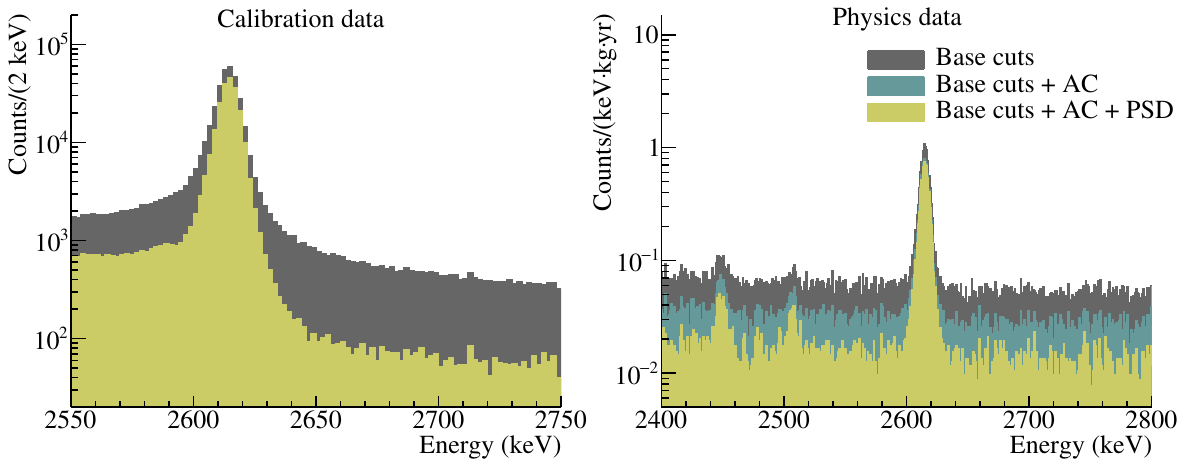}
  \caption{Pulse Shape Discrimination. Effect of the PSD cut on calibration data around the 2615\,keV line (left)
    and on physics data near \Qbb\ (right).
    In calibration data, the AC is not applied in order to maximize the statistics on the \G\ peaks,
    and the PSD mostly removes pileup events (events with more than one energy deposit in the time window).
    In physics data, the PSD mostly eliminates random noise events, which can correspond to either physical events with excessive noise or to noise-induced events with non-physical pulse shapes.
    Such events appear randomly across the energy spectrum,
    so the cut mostly acts on the continuum.
  }
  \label{fig:PSDcuts}
\end{figure}


\begin{figure*}[h]
  \centering
  \includegraphics[width=\textwidth]{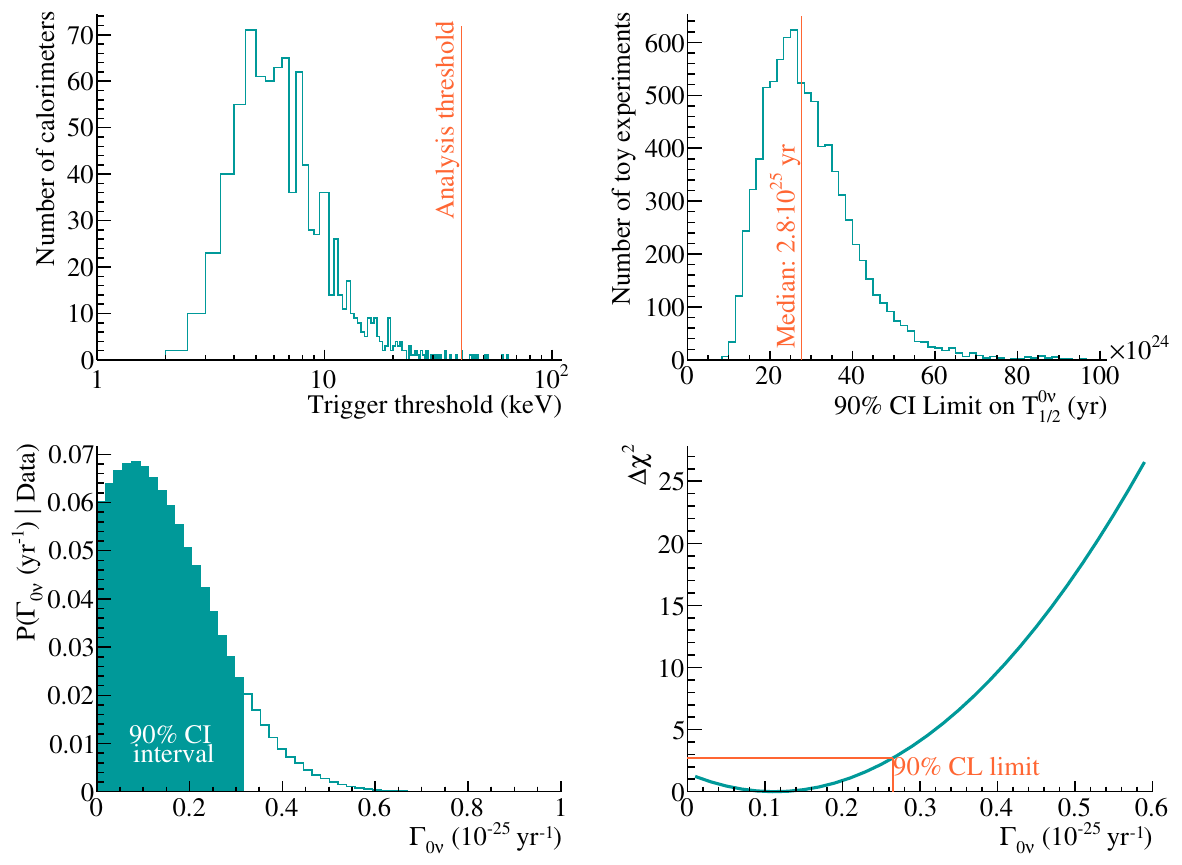}
  \caption{Optimum trigger and statistical analysis. Top left: Distribution of energy thresholds at 90\% trigger efficiency
    for the OT algorithm in a single dataset. The 40\,keV analysis threshold is indicated by the vertical line.
    Top right: 90\% CI exclusion limits on $T_{1/2}^{0\nu}$ from an ensemble of $10^4$ toy experiments generated
    with the background-only model, with background rates obtained from the background-only fit to the data.
    The median exclusion sensitivity is indicated by the orange line.
    Bottom left: Posterior probability distribution for $\Gamma_{0\nu}$ obtained from the Bayesian fit,
    with the 90\% CI highlighted.
    Bottom right: $\Delta\chi^2$ values obtained from the profile likelihood of $\Gamma_{0\nu}$,
    with $\Delta\chi^2=0$ being the most-favored value. The frequentist limit at 90\% confidence level (CL) is indicated.}
  \label{fig:OT_thresholds}\label{fig:DecayRate}\label{fig:Sensitivity}
\end{figure*}

\begin{figure}
  \centering
  \includegraphics[width=\textwidth]{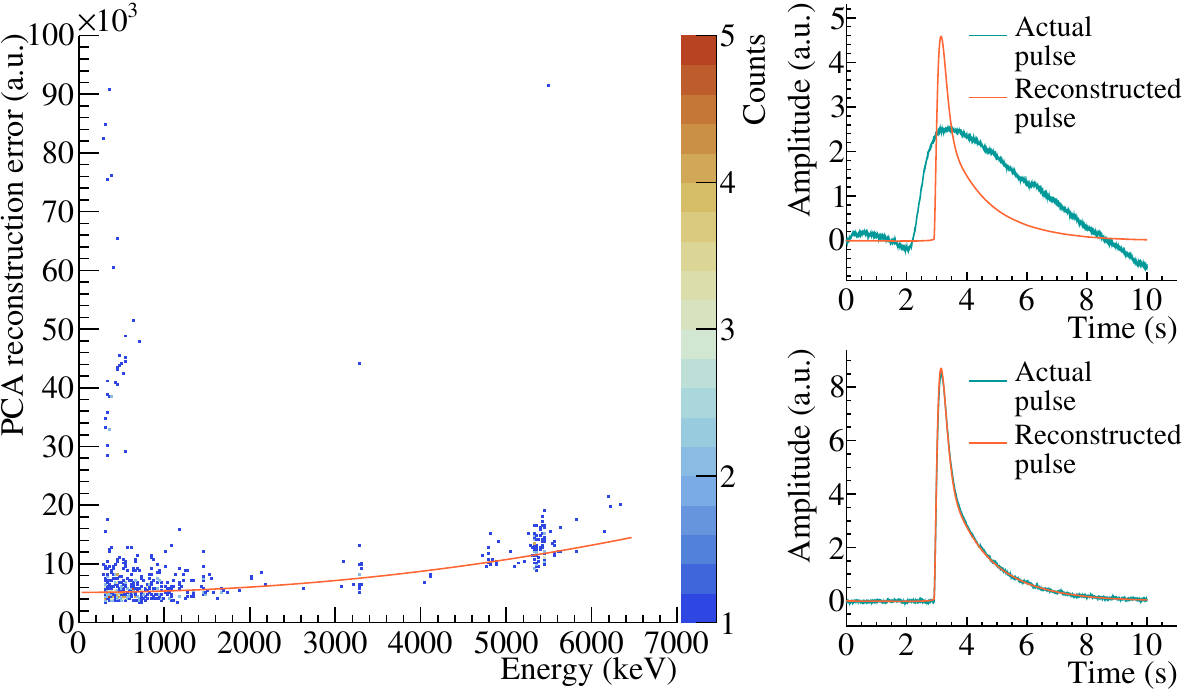}
  \caption{PCA performance. Left: example of a normalization fit of the PCA reconstruction error vs energy for a single calorimeter and dataset.
    The distribution contains only events that passed the other base cuts.
    The second order polynomial fit is shown in orange.
    Right: two example pulses from this calorimeter.
    The actual pulse is drawn in teal, and the corresponding reconstruction obtained by the PCA
    is drawn in orange. The top pulse deviates from the expected shape of a good pulse
    and is rejected, while the bottom one conforms to the expected response and is accepted.}
  \label{fig:pca_normfit}
\end{figure}

\begin{table}
  \centering
  \caption{Systematics affecting the \ovbb\ decay analysis.
    The total analysis efficiency is the product of all the efficiencies listed
    in Tab.~\ref{tab:Efficiencies} except containment.
    The PSD efficiency refers to its additional systematic uncertainty
    described in the text.
    The first four systematics are multiplicative effects and the impact of each is presented as a percentage.
    The last two systematics have a non-trivial effect on $\Gamma_{0\nu}$,
    hence we report the absolute effect.
    We report the variation induced on the marginalized 90\% CI limit (third column)
    and the posterior global mode $\hat{\Gamma}_{0 \nu}$ (last column).}\label{tab:Systematics}
  \includegraphics[width=.8\columnwidth]{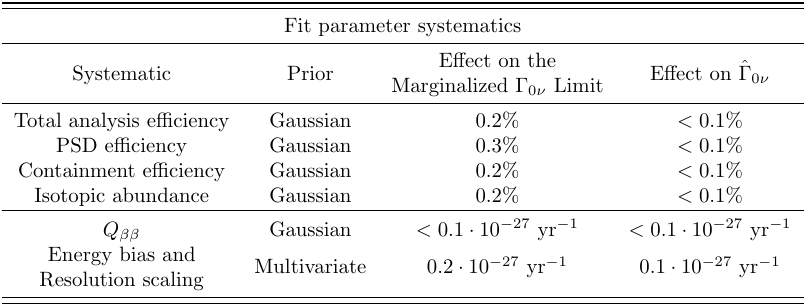}
\end{table}

\end{document}